\documentclass[pdflatex,sn-mathphys-num]{sn-jnl}

\usepackage{graphicx}%
\usepackage{rotating}%
\usepackage{natbib}%
\usepackage{multirow}%
\usepackage{amsmath,amssymb,amsfonts}%
\usepackage{amsthm}%
\usepackage{mathrsfs}%
\usepackage[title]{appendix}%
\usepackage{xcolor}%
\usepackage{textcomp}%
\usepackage{manyfoot}%
\usepackage{booktabs}%
\usepackage{algorithm}%
\usepackage{algorithmicx}%
\usepackage{algpseudocode}%
\usepackage{listings}%

\theoremstyle{thmstyleone}%
\theoremstyle{thmstyletwo}%

\theoremstyle{thmstylethree}%

\begin{document}

\title[Article Title]{A Data Science Approach to Calcutta High Court Judgments: An Efficient LLM and RAG-powered Framework for Summarization and Similar Cases Retrieval}


\author*[1]{\fnm{Puspendu} \sur{Banerjee}}\email{puspendu26th@gmail.com}

\author[1]{\fnm{Aritra} \sur{Mazumdar}}\email{aritra.raj2000@gmail.com}

\author[1]{\fnm{Wazib} \sur{Ansar}}\email{waakcs\_rs@caluniv.ac.in}

\author[2]{\fnm{Saptarsi} \sur{Goswami}}\email{sgakc@caluniv.ac.in}

\author[1]{\fnm{Amlan} \sur{Chakrabarti}}\email{acakcs@caluniv.ac.in}

\affil[1]{\orgdiv{A.K.Choudhury School of IT}, \orgname{University of Calcutta},  \city{Kolkata},  \state{West Bengal}, \country{India}}

\affil[2]{\orgdiv{Department of Computer Science}, \orgname{Bangabasi Morning College},  \city{Kolkata},  \state{West Bengal}, \country{India}}


\abstract{The judiciary, as one of democracy's three pillars, is dealing with a rising amount of legal issues, needing careful use of judicial resources. This research presents a complex framework that leverages Data Science methodologies, notably Large Language Models (LLM) and Retrieval-Augmented Generation (RAG) techniques, to improve the efficiency of analyzing Calcutta High Court verdicts.  Our framework focuses on two key aspects: first, the creation of a robust summarization mechanism that distills complex legal texts into concise and coherent summaries; and second, the development of an intelligent system for retrieving similar cases, which will assist legal professionals in research and decision making. By fine-tuning the Pegasus model using case head note summaries, we achieve significant improvements in the summarization of legal cases. Our two-step summarizing technique preserves crucial legal contexts, allowing for the production of a comprehensive vector database for RAG. The RAG-powered framework efficiently retrieves similar cases in response to user queries, offering thorough overviews and summaries. This technique not only improves legal research efficiency, but it also helps legal professionals and students easily acquire and grasp key legal information, benefiting the overall legal scenario.}




\maketitle

\section{Introduction}\label{sec1}

The legal domain is characterized by its vast and intricate nature, which includes a plethora of judgments, statutes, and legal documents. Each legal text can be lengthy and complex, making it challenging for legal professionals to swiftly incorporate and apply essential information. Natural Language Processing (NLP) approaches have advanced throughout time, revolutionizing how we process and analyze enormous amounts of information. These developments provide intriguing techniques for addressing the difficulties inherent in legal texts \citep{nallapati2016abstractive}. 

Summarization is an important NLP task that aims to condense a lengthy document into a brief and cohesive summary while  retaining its essential information. Summarization can be divided into two types: extractive and abstractive. Extractive summary selects and concatenates key lines from the original text, whereas abstractive summarization generates new sentences that encapsulate the fundamental concepts, frequently rephrasing the information in a more cohesive and concise manner \citep{zhong2020does}. Summarizing legal texts is particularly beneficial for several reasons. It can greatly decrease the time and effort necessary to comprehend long legal papers, speed up decision-making, and improve legal research by offering concise summaries of relevant cases \citep{filtz2020events}.

Despite these advancements, the legal profession faces significant challenges in efficiently processing and understanding extensive legal documents. Legal professionals frequently spend significant time reading through lengthy judgments and statutes to extract relevant information, which slows decision-making and reduces the overall efficiency of the legal system. 

Large Language Models (LLMs) trained on large datasets of legal texts have demonstrated potential in this area. However, LLMs have several limitations: they may not be fine-tuned particularly for legal contexts, may not be up to speed with the most recent legal data, may cause hallucinations, and may struggle to provide exact solutions to specific legal queries. Furthermore, general-purpose search engines frequently provide unorganized and overwhelming results, necessitating considerable effort to sort through irrelevant information. The absence of specialized tools for summarizing legal texts and retrieving similar cases worsen these issues, resulting in inefficiencies in legal research and analysis. There is an urgent need for innovative solutions that may give concise summaries and accurate retrieval of relevant cases to help legal practitioners with their work.

An LLM framework driven by Retrieval-Augmented Generation (RAG) can be used to overcome these constraints. This methodology leverages retrieval-based mechanisms along with the strengths of LLMs to guarantee contextually relevant and up-to-date replies. The accuracy and relevance of legal information retrieval are improved by RAG-powered frameworks, which gather relevant facts from a large database and produce logical answers \citep{lewis2020retrieval}.

The aim of this study is to make a substantial contribution to the field of legal research and analysis. Our framework gives detailed and organized summaries of court cases, in contrast to general-purpose search engines that typically provide unstructured and overwhelming results. With the use of case head note summaries, it fine-tunes the Pegasus model \citep{zhang2020pegasus} and significantly improves in legal text summarization. Our two-step summary procedure facilitates the development of an extensive vector database for RAG \citep{liu2019text} by maintaining important legal contexts. Furthermore, our RAG-powered structure facilitates the effective retrieval of relevant cases in response to user queries, offering an overview of case names, dates, results, citations, related acts articles sections invoked, and judgment summaries.

Additionally, we have created a comprehensive dataset of Calcutta High Court judgments by web scraping from indiankanoon.org \href{}. This dataset has been annotated using an LLM to overcome the time constraints associated with manual annotation, ensuring high-quality, consistent annotations and has been verified by legal experts.

In summary, the principal contributions of this paper are:
\begin{enumerate}
    \item Creating a dataset of Calcutta High Court judgments through web scraping and annotating it using an LLM, enhancing the quality and consistency of the dataset.
    \item Fine-tuning the Pegasus model with Case Head Note summaries to improve legal case summarization.
    \item Implementing a two-step summarization process to maintain the context of legal texts, crucial for building an accurate vector database for RAG.
    \item Developing a RAG-powered framework for retrieving similar cases, providing detailed overviews and summaries tailored to specific legal queries.
\end{enumerate}

By integrating these advancements, our paper addresses the shortcomings of existing LLMs and search engines, offering a robust and efficient tool for legal research and decision-making.

The remainder of the paper is structured as follows: Section \ref{sec2} delves into related work, positioning our work within the existing literature.  Section \ref{sec3} outlines the methodology that we used for this study. Section \ref{sec4} discusses our dataset and experimental setup in detail. Section \ref{sec5} presents insights derived from the experimental results. Finally, Section \ref{sec6} offers concluding remarks, along with the future scope of research in this domain.



\section{Related Works}\label{sec2}

Within this section, we provide a detailed overview of existing literature pertinent to our research topic. We seek to provide a contextual basis by reviewing earlier research, identifying essential themes, methodology, and findings that influence and add to the current study.

In order to facilitate Named Entity Recognition (NER) tasks, a new corpus of annotated legal named entities in Indian court decisions is introduced in the study by Kalamkar, Prathamesh, et al., 2022 \citep{kalamkar2022named}. About 46k annotated legal named entities corresponding to 14 legal entity kinds are included in the corpus.

The research by Mamakas, Dimitris, et al., 2022 \citep{mamakas2022processing} investigates how to handle lengthy legal documents more effectively by modifying pre-trained transformers like Longformer \citep{beltagy2020longformer} and LegalBERT \citep{chalkidis2020legal}. Using TF-IDF representations and eliminating duplicate sub-words during fine-tuning are two of the adjustments used to prevent going over the maximum input length limits.

A framework called CEEN, based on Reinforcement Learning (RL), is proposed by Lyu, Yougang, et al., 2022 \citep{lyu2022improving} to improve Legal Judgment Prediction (LJP). This methodology seeks to discriminate between fact descriptions by identifying unique criminal components and separating legal articles with comparable TF-IDF representations.

A corpus of English legal judgment papers is introduced, divided into logical sections, and annotated with predetermined rhetorical roles in the study by Kalamkar, Prathamesh, et al., 2022 \citep{kalamkar2022corpus}. To forecast rhetorical roles and enhance performance on tasks such as summarization and legal judgment prediction, a transformer-based baseline model is created.

Anand, D., \& Wagh, R. (2022). \citep{anand2022effective} suggest basic summarizing methods for Indian court judgment papers based on neural networks. Since the methods don't rely on manually created features or domain-specific expertise, they may be applied to other domains.

By utilizing event extraction from case facts,  Feng, Y., Li, C., \& Ng, V. (2022) \citep{feng2022legal} offers an Event-based Prediction Model (EPM) to enhance Legal Judgment Prediction (LJP). It presents a model that simultaneously learns LJP and event extraction, subject to limitations, and provides a hierarchical event structure for judicial situations.

Finally, Paul, Shounak, et al, 2023 \citep{paul2023pre} retrains well-known legal Pre-trained Language Models (PLMs) on Indian legal data and applies them to legal NLP tasks. This method improves performance on tasks such as Legal Statute Identification, Semantic Segmentation of Court Judgment Documents, and Court Appeal Judgment Prediction \citep{mazumdar2023efficient}.

\section{Methodology}\label{sec3}

In this section, the research methodology employed in the present study is outlined. The methodologies cover the overall design of the research and data analysis, providing a clear framework for understanding the research process.

\subsection{Data Collection and Annotation}\label{sec3:1}

\begin{figure}[htp]
    \centering
    \includegraphics[width=13 cm]{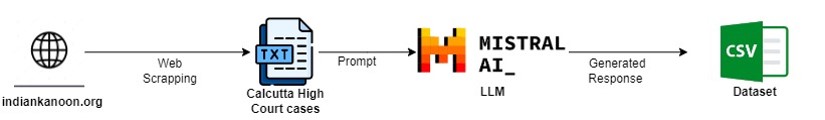}
    \caption{Data Collection and Annotation}
    \label{fig:data-collection-and-annotation}
\end{figure}

The process of gathering data for the approach starts with web scraping Calcutta High Court cases from indiankanoon.org \footnote{https://indiankanoon.org/} website. The dataset is built from this unprocessed legal data. Following prompting, the Mistral AI Large Language Model (LLM) processes these cases and produces structured replies that are saved in a CSV-formatted dataset, as shown in Fig. \ref{fig:data-collection-and-annotation}. This stage makes ensuring that the data is gathered, complete, and presented consistently so that it can be processed and analyzed further.

\subsection{Fine-tuning of Pegasus Model}\label{sec3:2}

\begin{figure}[htp]
    \centering
    \includegraphics[width=13 cm]{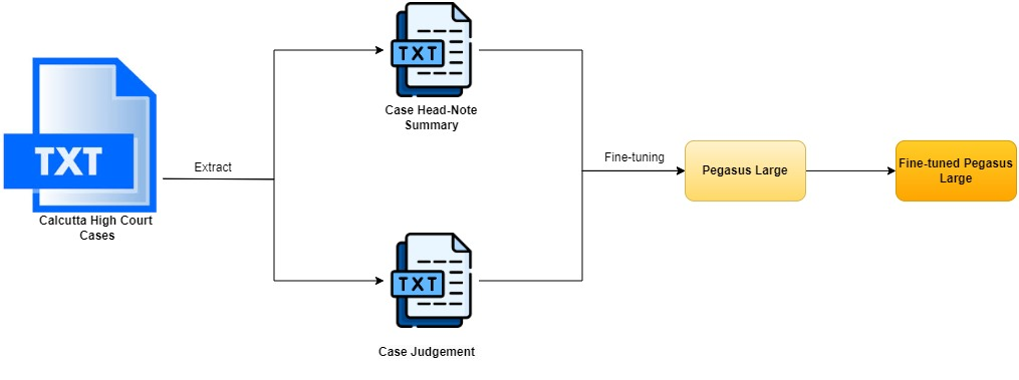}
    \caption{Fine-Tuning of Pegasus Model}
    \label{fig:fine-tuning-pegasus-model}
\end{figure}

The Pegasus model is fine-tuned using case headnote summaries taken from the Calcutta High Court cases in order to enhance the quality of summaries produced from the legal texts as shown in Fig. \ref{fig:fine-tuning-pegasus-model}. Case headnotes are perfect for training a summarization model because they are concise summaries that encapsulate the main points of court decisions. This process of fine-tuning improves the Pegasus model's capacity to provide more accurate and concise legal summaries that are suited to the legal field.

\subsection{Two-Step Summarization}\label{sec3:3}

\begin{figure}[htp]
    \centering
    \includegraphics[width=13 cm]{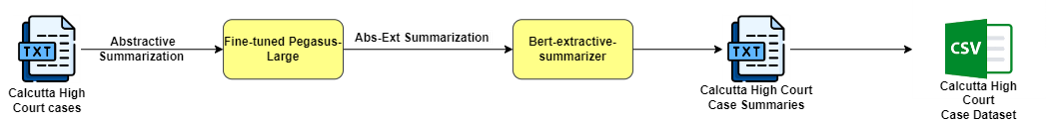}
    \caption{Two-step summarization}
    \label{fig:two-step-summarization}
\end{figure}

The two-step summarization process involves an initial abstractive step followed by an extractive step as demonstrated in Fig. \ref{fig:two-step-summarization}. The judgements are first fed into the fine-tuned Pegasus model, which generates a short and concise summary that captures the core ideas of the legal text. Key sentences are selected from the abstractive summary using the BERTSum model. This step ensures that the most relevant and important parts of the legal text are identified and form the summary.

\subsection{Embedding Generation and Vector Database Creation}\label{sec3:4}

\begin{figure}[htp]
    \centering
    \includegraphics[width=13 cm]{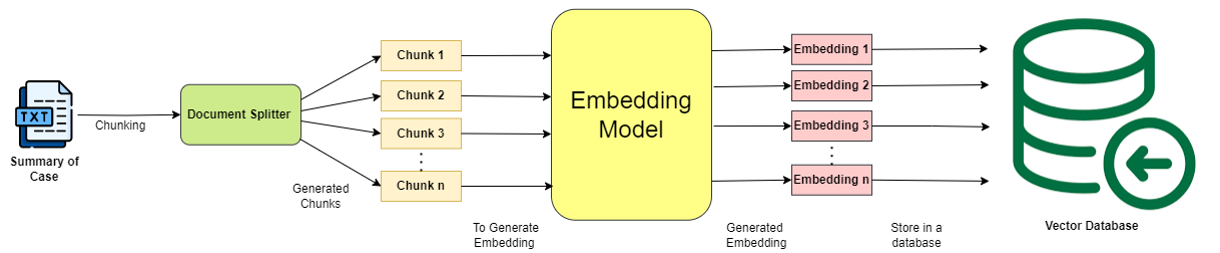}
    \caption{Embedding Generation and Vector Database Creation}
    \label{fig:embedding-generation-and-vector-db-generation}
\end{figure}

A document splitter subsequently divides the summaries — those generated using the two-step summarization process — into smaller, manageable chunks. We divided the text into 1024 token chunks for the vector database, with an overlap of 100 tokens. This method strikes a compromise between the requirements for comprehensive data and the embedding model's limitations. To create embeddings, each piece is then passed through an embedding model. These embeddings are stored in a vector database and serve as a representation of the summaries' semantic information. Fig. \ref{fig:embedding-generation-and-vector-db-generation} demonstrates this process. This is a crucial step since it converts the textual summaries into a format that makes comparison and querying efficient.

\subsection{RAG powered LLM Framework}\label{sec3:5}

\begin{figure}[htp]
    \centering
    \includegraphics[width=13 cm]{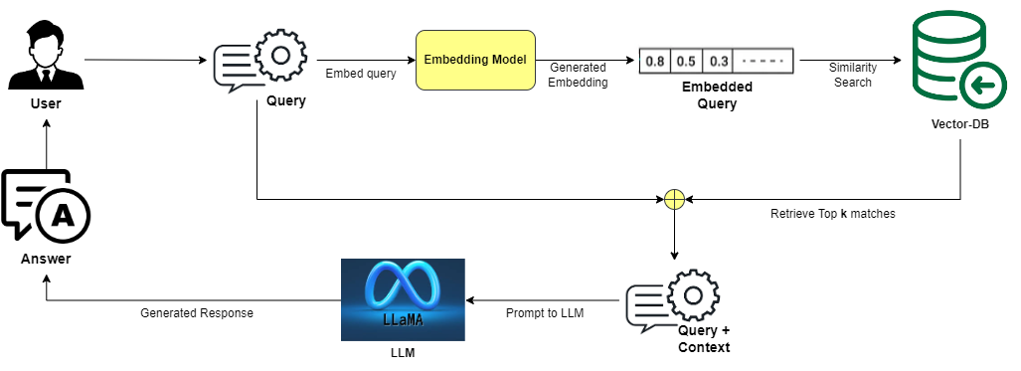}
    \caption{RAG powered LLM Framework}
    \label{fig:RAG-powered-LLM-Framework}
\end{figure}

The final step involves implementing the Retrieval-Augmented Generation (RAG) framework. When a user queries the system, the query is first embedded using the same embedding model that processed the summaries. These embedded queries are then used to perform similarity searches within the vector database, retrieving the top k matches that are most relevant to the query. The retrieved context, along with the original query, is then fed into the Llama-2 LLM. To ensure that the result is correct and contextually appropriate, the LLM provides a response based on the query as well as the obtained context as demonstrated in Fig. \ref{fig:RAG-powered-LLM-Framework}. By offering accurate and contextually rich answers, this method dramatically raises the efficacy and efficiency of legal research.

\section{Dataset and Experimental Setup}\label{sec4}

In this section, we provide comprehensive details regarding the data used in our study and outline the experimental setup employed for conducting the research.

\subsection{Dataset Information}\label{sec4:1}

\begin{table}[h]
\centering
\caption{Distribution of Case Types}
\label{tab:case-types}
\begin{tabular}{|l|r|}
\hline
\textbf{Case Type} & \textbf{Count} \\
\hline
Civil & 1155 \\
Criminal & 524 \\
Taxation & 436 \\
Contract & 412 \\
Constitutional & 130 \\
Labour Law & 123 \\
Others & 83 \\
Family Law & 33 \\
Probate & 19 \\
Workmen Compensation & 19 \\
Administrative Law & 15 \\
Election Law & 14 \\
Land Acquisition & 13 \\
Company Law & 8 \\
Industrial Disputes & 7 \\
Insolvency & 5 \\
Intellectual Property Rights & 5 \\
Preventive Detention & 5 \\
Banking & 4 \\
Citizenship & 4 \\
Trademark & 3 \\
Customs & 3 \\
Arbitration & 3 \\
Partition & 3 \\
Contempt of Court & 3 \\
Rent Control & 2 \\
Service Law & 2 \\
Bankruptcy & 2 \\
Quasi-Criminal & 1 \\
Insurance & 1 \\
\hline
\end{tabular}
\end{table}

\begin{figure}[htp]
    \centering
    \includegraphics[width=8 cm]{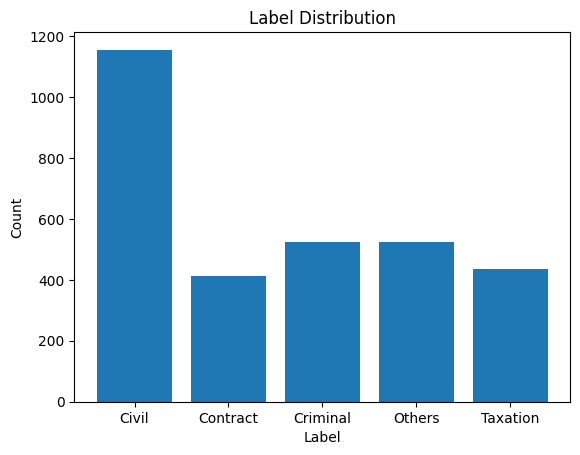}
    \caption{Distribution of cases in our dataset}
    \label{fig:case_type_plot}
\end{figure}

The Calcutta High Court Dataset is an extensive collection of legal judgments obtained through web scraping from indiankanoon.org website \footnote{https://indiankanoon.org/}. This dataset comprises approximately 130,000 raw text files of judgments. About 3,100 of these examples have undergone a thorough annotation process with the use of Mistral AI. The annotations offer a rich source of structured legal information, including case summaries and comprehensive metadata. Case Name, Date, Appellant, Respondent, Judges, Citations, Related Articles Sections and Acts, Case Type, Judgement, Summary, and Outcome of Appellant are among the annotated fields in the Calcutta High Court Dataset. Table \ref{tab:case-types} displays the various case types that have been annotated thus far, and Fig. \ref{fig:case_type_plot} provides a visual depiction of the same.

\subsection{Experimental Setup}\label{sec4:2}

\begin{table}[h]
\caption{Experimental Setup}\label{tab:exp-setup}%

    \begin{tabular}{|@{}l|l@{}|}
\hline
\textbf{Parameter} & \textbf{Value} \\
    \hline
RAM & 32GB \\
Accelerator Device & Tesla V100-PCIE GPU \\
Optimizer & AdamW \\
Learning Rate & 1.8e-5 \\
Weight decay & 0.2 \\
Max-chunk-size for summarization & at most 200 words (considering sentence boundaries) \\
Overlap-size for summarization & 2 sentences \\
Batch size & 3 \\
Epoch & 20 \\
Max-chunk-size for vector database document splitting & 1024 tokens \\
Overlap-size for vector database document splitting & 100 tokens \\
Number of Documents retrieved & 3 \\
\hline
\end{tabular}

\end{table}

Our experimental setup involved utilizing the Tesla V100-PCIE GPU with 32GB RAM as the accelerator device for running various machine learning models. We employed frameworks such as Langchain, PyTorch, TensorFlow, and scikit-learn for both training and evaluating results. Fine-tuning the Pegasus model involved using the AdamW optimizer with a learning rate of 1.8e-5, warm-up steps of 425, and a weight decay of 0.2. During fine-tuning, we used a batch size of 3 and conducted training on summarization tasks for 20 epochs. To address token size limitations in the Pegasus model, we chunked summaries based on sentence boundaries, ensuring each chunk contained a maximum of 200 words with an overlap of two sentences to maintain context. Similarly, for the vector database, we split text into chunks of 1024 tokens with an overlap of 100 tokens. This approach aimed to balance the need for detailed information with the constraints of the embedding model. Additionally, the RAG framework was utilized to retrieve the top 3 documents through similarity search and generate outputs based on this retrieved information. Table \ref{tab:exp-setup} gives a brief overview of our experimental setup.

\section{Results}\label{sec5}

In this section, we present an ablation study comparing various summarization techniques: Extractive Summarization, Abstractive Summarization (with and without fine-tuning), Ext-Abs Summarization (with and without fine-tuning), and Abs-Ext Summarization (with and without fine-tuning). We evaluate their performance using precision, recall, and F1 score metrics with ROUGE-1, ROUGE-2, and ROUGE-L as shown in Tables~\ref{tab:recall}, \ref{tab:precision}, and \ref{tab:f1}.

\begin{table}[h]
\caption{Evaluation of different summarization techniques on Recall metric}
\label{tab:recall}
\centering
\begin{tabular}{|l|c|c|c|}
\hline
\textbf{Summarization Technique} & \textbf{ROUGE-1} & \textbf{ROUGE-2} & \textbf{ROUGE-L} \\
\hline
Extractive Summarization & \textbf{0.82} & 0.50 & 0.44 \\
\hline
Abstractive Summarization (without FT) & 0.69 & 0.39 & 0.37 \\
Abstractive Summarization (with FT) & \textbf{0.80} & \textbf{0.50} & \textbf{0.47} \\
\hline
Ext-Abs Summarization (without FT) & 0.41 & 0.21 & 0.21 \\
Ext-Abs Summarization (with FT) & 0.54 & 0.27 & 0.28 \\
\hline
Abs-Ext Summarization (without FT) & 0.41 & 0.21 & 0.22 \\
Abs-Ext Summarization (with FT) & 0.58 & 0.31 & 0.30 \\
\hline
\end{tabular}
\end{table}

\begin{table}[h]
\caption{Evaluation of different summarization techniques on Precision metric}
\label{tab:precision}
\centering
\begin{tabular}{|l|c|c|c|}
\hline
\textbf{Summarization Technique} & \textbf{ROUGE-1} & \textbf{ROUGE-2} & \textbf{ROUGE-L} \\
\hline
Extractive Summarization & 0.42 & 0.26 & 0.22 \\
\hline
Abstractive Summarization (without FT) & 0.53 & 0.30 & 0.28 \\
Abstractive Summarization (with FT) & 0.39 & 0.25 & 0.23 \\
\hline
Ext-Abs Summarization (without FT) & 0.58 & 0.30 & 0.31 \\
Ext-Abs Summarization (with FT) & 0.69 & 0.37 & 0.38 \\
\hline
Abs-Ext Summarization (without FT) & 0.62 & 0.35 & 0.34 \\
Abs-Ext Summarization (with FT) & \textbf{0.69} & \textbf{0.37} & \textbf{0.38} \\
\hline
\end{tabular}
\end{table}

\begin{table}[h]
\caption{Evaluation of different summarization techniques on F1-score metric}
\label{tab:f1}
\centering
\begin{tabular}{|l|c|c|c|}
\hline
\textbf{Summarization Technique} & \textbf{ROUGE-1} & \textbf{ROUGE-2} & \textbf{ROUGE-L} \\
\hline
Extractive Summarization & 0.53 & \textbf{0.32} & 0.28 \\
\hline
Abstractive Summarization (without FT) & 0.56 & 0.32 & 0.29 \\
Abstractive Summarization (with FT) & 0.50 & 0.31 & 0.29 \\
\hline
Ext-Abs Summarization (without FT) & 0.47 & 0.24 & 0.25 \\
Ext-Abs Summarization (with FT) & 0.52 & 0.26 & 0.27 \\
\hline
Abs-Ext Summarization (without FT) & 0.47 & 0.25 & 0.22 \\
Abs-Ext Summarization (with FT) & \textbf{0.56} & 0.30 & \textbf{0.29} \\
\hline
\end{tabular}
\end{table}

Extractive summarization shows higher F1 scores across all ROUGE metrics compared to abstractive summarization when applied to legal documents. Precision is also higher in extractive summarization for ROUGE-1, though abstractive summarization has better recall scores. The abstractive Pegasus model excels in recall, suggesting it captures more relevant information from the source text but might generate more irrelevant content, reducing precision and F1 scores.

Abs-Ext summarization consistently outperforms Ext-Abs summarization across all metrics (F1, precision, and recall) in the context of legal documents. The Abs-Ext model demonstrates a balance between precision and recall, suggesting that this approach is more effective in generating coherent and relevant summaries. The Ext-Abs model, while still effective, lags behind slightly, indicating that the ordering of extraction and abstraction processes influences the summarization quality.

\section{Conclusion and Future Works}\label{sec6}

Our study effectively addresses the challenges posed by the increasing volume and complexity of legal texts through two-step summarization and efficient case retrieval. By developing a RAG-powered framework, we ensure precise and contextually relevant retrieval of similar cases, significantly improving the efficiency of legal research. The creation of a robust, LLM-annotated dataset of Calcutta High Court judgments ensures consistent and accurate annotations, forming a solid foundation for our system. This work not only streamlines legal research for professionals but also provides a valuable educational tool for law students and budding professionals. Overall, our study demonstrates the transformative potential of integrating data science with legal research to enhance decision-making and efficiency in the legal field.

One promising avenue for future development is the integration of reinforcement learning with human feedback. This approach will enable our LLM to learn from real-world legal expertise, thereby improving its ability to generate more accurate and contextually appropriate summaries and responses. This can greatly improve the dependability and performance of the model. There is also considerable potential to expand the framework to other jurisdictions, both within India and internationally. This expansion would involve creating similar datasets and fine-tuning models for different legal systems and languages, thereby broadening the applicability and utility of our system. To ensure the model remains current and relevant, regular updates to the LLM with the latest legal texts and judgments will be essential. This will keep the model up-to-date with the most recent legal developments and trends, ensuring its continued usefulness and accuracy. Exploring advanced annotation techniques, including semi-supervised and unsupervised methods, could further improve the quality and scalability of our dataset creation process. 

By pursuing these future directions, we aim to create a more robust, versatile, and user-friendly legal information system that significantly enhances the efficiency and accuracy of legal research and decision-making.


\bibliography{sn-bibliography}

\end{document}